\documentclass[aps,prl,floats,
  twocolumn,showpacs,superscriptaddress,reprint]{revtex4}
\usepackage{graphicx,epsfig}
\usepackage{graphics,dcolumn,bm,epic, eepic,fleqn,float}
\usepackage{amssymb,amsmath,amsfonts,multirow,rotate,color}
\usepackage{soul}
\usepackage{wasysym}
\usepackage{subfigure}

\definecolor{Myorange}{cmyk}{0,0.42,1,0}
\definecolor{MyBrick}{rgb}{0.84,0.01,0.01}

\newcommand{\avg}[1]{\langle #1 \rangle}

\newcommand{\lay}[1]{^{[#1]}}

\begin{document}

\title{Efficient exploration of multiplex networks}

\author{Federico Battiston}
\affiliation{School of Mathematical Sciences, Queen Mary University of
  London, Mile End Road, E1 4NS, London (UK)}
\author{Vincenzo Nicosia}
\affiliation{School of Mathematical Sciences, Queen Mary University of
  London, Mile End Road, E1 4NS, London (UK)}
\author{Vito Latora}
\affiliation{School of Mathematical Sciences, Queen Mary University of
  London, Mile End Road, E1 4NS, London (UK)}
\affiliation{Dipartimento di Fisica ed Astronomia, Universit\`a di Catania and INFN, I-95123 Catania, Italy}

\begin{abstract}
Efficient techniques to navigate networks with local
  information are fundamental to sample large-scale online social
  systems and to retrieve resources in peer-to-peer systems. Biased
  random walks, i.e. walks whose motion is biased on properties of
  neighbouring nodes, have been largely exploited to design smart
  local strategies to explore a network, for instance by constructing
  maximally mixing trajectories or by allowing an almost uniform
  sampling of the nodes. Here we introduce and study biased random
  walks on multiplex networks, graphs where the nodes are related
  through different types of links organised in distinct and
  interacting layers, and we provide analytical solutions for their
  long-time properties, including the stationary occupation
  probability distribution and the entropy rate. We focus on
  degree-biased random walks and distinguish between two classes of
  walks, namely those whose transition probability depends on a number
  of parameters which is extensive in the number of layers, and those
  whose motion depends on intrinsically multiplex properties of the
  neighbouring nodes. We analyse the effect of the structure of the
  multiplex network on the steady-state behaviour of the walkers, and
  we find that heterogeneous degree distributions as well as the
  presence of inter-layer degree correlations and edge overlap
  determine the extent to which a multiplex can be efficiently
  explored by a biased walk. Finally we show that, in real-world
  multiplex transportation networks, the trade-off between efficient
  navigation and resilience to link failure has resulted into systems
  whose diffusion properties are qualitatively different from those of
  appropriately randomised multiplex graphs. This fact suggests that
  multiplexity is an important ingredient to include in the modelling
  of real-world systems.
\end{abstract}


\maketitle

The network paradigm has proven to be a successful framework to study
the intricate patterns of relations among the constituents of
real-world complex systems, from the Internet to the human
brain~\cite{Newman2003rev,Boccaletti2006}, and has revealed
that the dynamical behaviours observed in such systems, such as
information spreading, diffusion, opinion formation and
synchronisation, are quite often affected ---and to some extent
determined--- by the structure of the underlying interaction
network~\cite{Pastor2001,Arenas2008,Castellano2009,Bullmore2009}. However, the
recent availability of massive data sets of social, technological and
biological systems has suggested that the classical complex network
approach might fall somehow short in modelling systems whose
elementary units can interact through more than one type of
connections. This is typical of many real-world systems, such as social networks, where people are connected through a variety of social
relationships, e.g. kinship, friendship, collaboration, competition, or transportation systems, which often exploit different communication channels~\cite{Szell2010,Cardillo2013,Gallotti2014,Battiston2016}. 
Such systems can be treated in terms of
\emph{multi-layer} or \emph{multiplex networks}~\cite{DeDomenico2013,Battiston2014,Boccaletti2014,Kivela2014}, where each layer describes a particular type of interaction among the nodes of the graph. 
Some recent works have confirmed that multi-layer networks are
characterised by new levels of
complexity~\cite{DeDomenico2015} and
that the interaction of multiple network layers can produce new interesting
 dynamical
behaviours~\cite{Baxter2012, Gomez-Gardenes2012, Gomez2013, Cozzo2013, Nicosia2014sync,Diakonova16}.

In the realm of dynamical processes on networks~\cite{Barrat2008} the
simplicity and -still- the richness of random walks has attracted much
attention in recent years~\cite{Noh2004, Yang2005}. Random
  walks are the most simple way to explore a network using only local
  information, and the steady-state properties of a walk, including
  characteristic times, limiting occupation probability, and coverage,
  have tight relationships with the structure of the graph upon which
  the walk takes place~\cite{zhang2013,yuanlin-zhang2013}. For this
  reason, random walks have also been successfully used as probes of
  network properties, with applications ranging from community
  detection~\cite{Zhou2003,Rosvall2008,Fallani2014} to taxonomy of
  real-world networks~\cite{Nicosia2014Exponents}. Moreover, specific
  flavours of random walks are widely used for the exploration of
  online social networks, information networks and the like.

A class of random walkers of particular interest is that of walkers
whose motion is biased on the structural properties of the
network~\cite{Gomez-Gardenes2008}. In its simplest possible version,
the considered \emph{biased random walks} are Markov processes whose
transition probability is a parametric function of the topological
properties of the destination node. In this way, by tuning the
parameters of the biasing function one can force the walk to
preferentially visit, or avoid, nodes exhibiting high or low values of
given topological descriptors, such as the degree, clustering or
betweenness. In particular, degree-biased random walks have
  been used to define new centrality measures~\cite{Lee2009,
    Delvenne2011}, identify communities \cite{Zlatic2010}, and provide
  optimal exploration of a network using only local
  information~\cite{Sinatra2011}. It has also been found that the
  dynamics of degree-biased random walks is strongly affected by the
  presence of degree-degree correlations in the structure of the
  network~\cite{Fronczak2009, Baronchelli2010, Bonaventura2014}, so
  that an appropriate choice of the structural bias can be used to
  perform efficient sampling of unknown networks.  

In this Article we study several ways in which random walks
  can be extended to multi-layer networks, and we show how to devise
  appropriate ways to bias the walkers on the topological properties
  of the nodes at each layer in order to perform an efficient
  exploration of such systems. We notice that random walks have
  already been applied to multi-layer networks, e.g. to quantify the
  impact of failures in interconnected
  systems~\cite{DeDomenico2014PNAS}.  However, we will focus here on
  biased random walks and will investigate how the biasing function
  affects the dispersiveness of the walk and the steady-state
  occupation probability distribution. The aim is to find walks which
  visit far away regions of a multiplex network within a relatively
  small number of steps, a property related to the dispersiveness of
  the walk, and, at the same time, guarantee that the probability for
  a walker to visit any node in the system is as close as possible to
  uniform, thus allowing to sample unknown graphs in an efficient
  way.

The presence of many interdependent layers allows to construct
  several classes of biased random walks, and in particular what we
  call \textit{extensive walks} and \textit{intensive walks}, where
  the difference between the two classes is in the dependence of the
  parameters of the biasing function on the number of layers of the
  system. In the former case, the biasing function depends on the
  structural properties of the destination node at all the layers of
  the system (thus, the number of parameters is extensive in the
  number of layers), while in the latter case the bias depends on
  intrinsically multiplex properties of the destination node, which do
  not depend explicitly on the number of layers of the network.


For both classes of biasing functions, we provide analytical
  closed forms for the long-time properties of the random walks, in
  terms of stationary probability distribution and entropy
  rate~\cite{Cover1991}, and we study the effect of different
  structural properties, including the number of layers, the presence
  and sign of inter-layer degree correlations, the redundancy of edges
  across layers, the density of the multiplex and the heterogeneity of
  the degree distributions, on the steady-state behaviour of these
  walks. We find that all these properties have a remarkable effect on
  the maximal dispersiveness and on the steady-state occupation
  probability of biased random walks.

Finally, we study the diffusion properties of several
  real-world multiplex networks, namely the six continental airline
  transportation networks, and we show that in those cases the
  pressure to provide robust route alternatives has somehow hindered
  the overall diffusion properties of those systems.

\section{General features of biased random walks}

Let us consider a $M$-layer multiplex network of $N$ nodes, i.e. a
multi-layer graph in which each node can interact with the other ones
by means of $M$ different kinds of relationships. A multiplex is fully
described by the $M$-dimensional array of the adjacency matrices of
its layers $\mathcal{A}=\{A\lay{1}, A\lay{2}, \ldots, A\lay{M}\}$,
where $A\lay{\alpha}=\{a\lay{\alpha}_{ij}\}\in\mathbb{R}^{N\times N}$
and $a\lay{\alpha}_{ij}=1$ if node $i$ and node $j$ are connected at
layer $\alpha$. In the following we assume the layers to be unweighted, but all the results can be easily extended to to the case of weighted multiplexes.

In general, a random walker on a multiplex is not constrained on a
single layer and can exploit all the connections pointing out of the
current node, at all layers.  A synthetic --yet incomplete--
description of the topology of a multiplex is provided by the
overlapping adjacency matrix $\mathcal{O}=o_{ij}$, whose entries
$o_{ij}= \sum_{\alpha}a_{ij}^{[\alpha]}$ account for the total number
of connections between two nodes across all
layers~\cite{Battiston2014}. In particular, we consider the class of
Markovian random walks defined by the transition probabilities:
\begin{equation}
  \pi_{ji}=\frac{o_{ij}f_j}{\sum_j o_{ij}f_j}.
  \label{eq:walk}
\end{equation}
This set up is very general and allows for a variety of different
motion rules. In fact, $f_j$ can be either a function of some
topological multiplex properties of the arrival node $j$, or an
informative combination of some structural features of the destination
node, measured at all or at a fraction of the layers.  Notice that the
unbiased random walk on the multiplex is obtained by setting $f_j=1,
\> \forall j\in V$. In this case a walker jumps out of node $i$ by
traversing one of the edges incident on $i$ chosen with uniform
probability and independently on the layer to which it belongs.
It is worth noting that the use of the overlapping adjacency
  matrix $\{o_{ij}\}$ does not automatically make the walk in
  Eq.~(\ref{eq:walk}) equivalent to a random walk on the aggregated
  graph obtained by flattening all the layers in a single network. In
  general, if the biasing function $f_{j}$ depends, either explicitly
  or implicitly, on the structural properties of node $j$ in the
  multiplex network, the walk in Eq.~(\ref{eq:walk}) cannot be
  directly mapped on an equivalent walk on the aggregated graph.


\textit{Stationary probability distribution. ---} Starting from the
one-step transition probability given in Eq.~\ref{eq:walk} we derive
closed forms for several asymptotic properties of the walk.
Following an approach similar to that used in
  Ref.~\cite{Gomez-Gardenes2008}, we now show that for any choice of
the biasing function $f_j$ the stationary probability distribution
$\bm{p}^*=\{p^*_i\}$ of biased walks on multiplex networks can be
analytically derived, under the hypotheses that \textit{i)} the
topological overlapping matrix $\mathcal O$ is primitive and that
\textit{ii)} $f_j$ is a time-invariant function of any property of the
destination node $j$.
We start by considering the probability $p_{i \to j}(t)$ that a walker
starting at node $i$ will be found on node $j$ after exactly $t$ time
steps:
\begin{equation}
  p_{i \to j}(t)=\sum_{j_1,j_2,\ldots,j_{t-1}}\!\!\!\!\pi_{j_1,i}\times
  \pi_{j_2,j_1}\times\ldots \times \pi_{j,j_{t-1}},
  \label{eq:pij}
\end{equation}
and the dual probability $p_{j\to i }(t)$:
\begin{equation}
  p_{j \to i}(t)=\sum_{j_1,j_2,\ldots,j_{t-1}}\!\!\!\!\pi_{j_1,j}\times
  \pi_{j_2,j_1}\times\ldots \times \pi_{i,j_{t-1}}.
  \label{eq:pji}
\end{equation}
Comparing Eq.~(\ref{eq:pij}) with Eq.~(\ref{eq:pji}) and considering
that the multiplex is undirected (i.e., $o_{ij}=o_{ji}$), we obtain
\begin{equation}
c_i f_i  p_{i \to j}(t) = c_j f_j  p_{j \to i}(t), \quad \forall
i,j\in V
\end{equation}
where $c_i=\sum_j o_{ij}f_j$. If the matrix $\mathcal{O}$ is
primitive, then a stationary probability distribution exists and
$\lim_{t\to \infty} p_{i\to j}(t) = p^*_j$, leading to the expression:
\begin{equation}
  c_i f_i p^*_j = c_j f_j p^*_i.
\end{equation}
By imposing the normalisation condition $\sum_j p^*_j=1$ we finally
get:
\begin{equation}
  p_i^*=\frac{c_i f_i}{\sum_{\ell} c_{\ell} f_{\ell}}.
  \label{eq:stationary}
\end{equation}
We notice that Eq.~(\ref{eq:stationary}) is quite general, since it
does not explicitly depend on the form of the biasing function or on
the actual structure of each layer or of the topological overlapping
matrix $\mathcal{O}$. 

In many real-world application scenarios, e.g. in crawling the
  structure of online social networks, it is important to guarantee
  that for long enough times the walk will end up visiting all the
  nodes of the graph with the same probability. It is easy to prove
  that an unbiased random walk is not a good choice in this case,
  since its steady-state occupation probability distribution is
  proportional to the degree sequence, hence an appropriate bias
  should be used to avoid to visit hubs more frequently than
  poorly-connected nodes. In practice, it is not always possible to
find a walk which produces exactly the same stationary occupation
probability distribution for all the nodes, i.e.  $p^*_i=\overline{p}
= 1/N,\> \forall i$. However, one could instead require that the
resulting stationary probability distribution, although not equal for
all nodes, has the minimum possible variance. In particular, in the
following we will focus on the normalised standard deviation of the
stationary probability distribution:
\begin{equation}
  \eta(p^*) = \frac{\sigma(p^*)}{\mu(p^*)}
\end{equation}
where $\mu(p^*)$ and $\sigma(p^*)$ are the average and the standard
deviation of $\bm{p}^*$, respectively. We will look for suitable
combinations of the parameters of the walk that produce the smallest
possible value of $\eta(p^*)$, corresponding to the maximum uniformity
of the accessibility of the nodes attainable on a certain multiplex
network.

\textit{Entropy rate. ---} One classical measure to quantify
  the mixedness or dispersiveness of a walk on a graph is the entropy
rate $h=\lim_{t \to \infty} S_t/t$~\cite{Cover1991}, where $S_t$ is
the Shannon entropy of the set of all the trajectories of length $t$
generated from the walk rule, and $h$ is the minimum amount of
information necessary to describe the process~\cite{Cover1991}. In
particular, $h=0$ only if the walk generates exactly one possible
trajectory, while $h$ is maximum when all the trajectories are
equiprobable. Intuitively, walks with a high mixedness can
  explore remote regions of a graph within a relatively small number
  of time-steps. This property is again desirable for the efficient
  exploration of unknown networks, where only local information is
  available. In particular, it is interesting to find a biasing
  function which guarantees that the walk does not remain trapped for
  too long in any region of the graph, and this is usually obtained by
  maximising the dispersiveness of the walk.

It is possible to show that the entropy rate for a Markov process can
be expressed as
\begin{equation}
h=-\sum_{i,j}\pi_{ji}p_i^*\ln(\pi_{ji}),
\label{eqentropyrate}
\end{equation}
which means that $h$ depends only on the walk rule
$\pi_{ij}$ and on the stationary probability
distribution~\cite{Gomez-Gardenes2008}. By substituting the analytical
expression for ${\bm p^*}$ given in Eq.~(\ref{eq:stationary}) into
Eq.~(\ref{eqentropyrate}) we get:
\begin{equation}
h = - \frac{1}{\sum_i c_i f_i}\left[\sum_i f_i \sum_j o_{ij} f_j
  \ln(o_{ij}f_j)-\sum_i f_i c_i \ln(c_i) \right].
\end{equation}
This expression has a natural upper bound, which reflects the case of
random walks where all trajectories of the same length have equal
probability. It is interesting to notice that, as shown by Burda et
al. in Ref.~\cite{Burda2009}, the maximal value of entropy rate
attainable by any walk on a given single-layer graph depends on the
structure of the graph, and in particular for an undirected graph it
is equal to $\ln \lambda_{\rm max}$, where $\lambda_{\rm max}$ is the
maximum eigenvalue of the adjacency matrix of the graph.

This result can be extended to the case of walks on multiplex networks
as follows.  The total number of trajectories of length $t$ generated
by a walk defined as in Eq.~(\ref{eq:walk}) is equal to
$N_t=\sum_{i,j}(\mathcal O^t)_{ij}$, where $\mathcal O^t$ is the
$t$-th power of the overlapping adjacency matrix. In the limit of
large $t$, we have
\begin{equation}
\widetilde{h}_{\rm max}=\lim_{t \to \infty} \frac{\ln N_t}{t}=\ln
\lambda_{\rm max},
\label{eq_maxentrmultiplex}
\end{equation} 
where $\lambda_{\rm max}$ is now the maximum eigenvalue of the
overlapping adjacency matrix $\mathcal O$ (this result is a direct
consequence of the application of the power method). In general, the
maximal value of the entropy rate attainable with a particular motion
rule will be smaller than or at most equal to $\widetilde{h}_{\rm
  max}$. Since obtaining high mixedness is a desirable property of a
walk in many real-world applications, such as when searching for a
given resource on a graph, in the following we will look for
combinations of the parameters of different motion rules which can
produce high values of $h$, to better approximate the corresponding
value of $\widetilde{h}_{\rm max}$ allowed by the structure of the
network.

\textit{Heterogeneous mean-field. ---} In the particular case in which
the bias function $f_i$ depends only on the (vectorial) degree
$\bm{k}_i=\{k\lay{1}_i, k\lay{2}_i, \ldots, k\lay{M}_i\}$ of node $i$,
where by definition $k\lay{\alpha}_i=\sum_{j}a\lay{\alpha}_{ij}$ is
the degree of node $i$ at layer $\alpha$, the expression for the
stationary probability distribution can be considerably
simplified. Let us consider a heterogeneous mean-field, in which all
the nodes belonging to the same degree class $\bm{k}$ are structurally
indistinguishable. Under these assumption, and since $f_i$ depends
only on the degree, then for all the nodes $i$ having the same degree
$\bm{k}_i = \bm{k}$ we have $f_i = f_{\bm{k}_i} = f_{\bm{k}}$, but
also $c_{i} = c_{\bm{k}_i} = c_{\bm{k}}$, and similarly:
\begin{equation}
  p^{*}_{\bm{k}} =\frac{1}{C} f_{\bm{k}} c_{\bm{k}} =
  \frac{1}{C} f_{\bm{k}}\sum\limits_{\bm{k'}}
  o_{\bm{k}\bm{k'}}f_{\bm{k'}}
  \label{eq:meanfield}
\end{equation}
where $C$ is an appropriate normalisation constant to ensure that
$\sum_{\bm{k}}p^{*}_{\bm{k}}=1$. Eq.~(\ref{eq:meanfield}) means that
all the nodes in the same degree class will have the same steady-state
probability of being visited by the walk. Notice that $o_{\bm{k k'}}$
is the expected number of edges connecting two nodes whose multiplex
degree is respectively equal to $\bm{k}$ and to $\bm{k'}$. If we
assume that there are no edge correlations, i.e. that the probability
of having $a\lay{\alpha}_{ij}=1$ does not depend on the probability of
having $a\lay{\beta}_{ij}=1$ for all the possible $\beta\neq \alpha$,
then we can write:
\begin{equation}
  p^{*}_{\bm{k}} = \frac{1}{C} f_{\bm{k}} \sum_{\bm{k'}}f_{\bm{k'}}
  \sum_{\alpha=1}^{M} k\lay{\alpha} P({k'}\lay{\alpha}|k\lay{\alpha})
\end{equation}
since the expected number $o_{\bm{kk'}}$ of edges between a node with
degree $\bm{k}$ and a node with degree $\bm{k'}$ is actually equal to
the sum of the expected number of edges connecting these two nodes at
each of the $M$ layers (we indicate by ${k'}\lay{\alpha}$ the degree
at layer $\alpha$ of a node whose vectorial degree is equal to
$\bm{k'})$. If we additionally assume that there are no
  intra-layer correlations, then:
\begin{equation}
  P({k'}\lay{\alpha}|k\lay{\alpha}) = q_{{k'}\lay{\alpha}} =
  \frac{{k'}\lay{\alpha} P({k'}\lay{\alpha})}{\avg{{k'}\lay{\alpha}}}
\end{equation}
where $P({k'}\lay{\alpha})$ is the degree distribution at layer
$\alpha$. In the end we find:
\begin{equation}
  p^{*}_{\bm{k}} = \frac{1}{C}
  f_{\bm{k}}\sum_{\bm{k'}}f_{\bm{k'}}\sum_{\alpha=1}^{M}
  \frac{k\lay{\alpha}{k'}\lay{\alpha}P({k'}\lay{\alpha})}{\avg{{k'}\lay{\alpha}}}.
  \label{eq:p_star_k}
\end{equation}
This expression for $p^*_{\bm{k}}$ is quite general, and in
  particular it is valid even in the presence of inter-layer
  degree-correlations~\cite{Nicosia2014corr}.  Since the heterogeneous
  mean-field discards intra-layer and edge correlations, which usually
  contribute to hinder the dispersiveness of a walk,
  Eq.~(\ref{eq:p_star_k}) can be readily plugged into the expression
  of the entropy rate in Eq.~(\ref{eqentropyrate}) to obtain an
  estimate of the maximum value of $h$ attainable with a given biasing
  function on a multiplex network with an assigned multiplex degree
  sequence $\{\bm{k}_i\}$.

\section{Classes of biased random walks}
The introduction of a biasing function in the motion rule is mainly
motivated by the necessity to obtain an exploration of the graph which
is more efficient, i.e., faster with respect to the time needed to
visit all the nodes, or more homogeneous, i.e., avoiding
heterogeneities in the stationary distribution probability, in order
to explore with the same probability each node of the graph. In single
layer networks these two aims are in general antithetical. For
instance, a biasing function which maximises the mixing of the walk
(corresponding to higher values of entropy rate) usually produces a
quite heterogeneous stationary occupation probability, mainly due to
the fact that a better mixing is obtained by exploiting the central
role played by hubs. High values of $h$ are usually achieved in a
single-layer uncorrelated graph by a degree-biased walk $\pi_{ji} \sim
k_j^{b}$ with $b=1$, and in general with a bias $b>0$ in graphs with
non-trivial degree-degree correlations~\cite{Gomez-Gardenes2008}. On
the other hand, a uniform stationary occupation probability is
obtained by using $\pi_{ji} \sim k_j^{b}$ with $b=-1$ in uncorrelated
graphs, and in general by a value of $b<0$ for graphs with
degree-degree correlations, which corresponds to forcing the walkers
to preferentially move towards poorly connected
nodes~\cite{Bonaventura2014}.

The richness of multi-layer networks allows the exploration of more
complex biasing functions and, as we will show in the following,
usually produces quite interesting dynamics. The reason of such
richness is that the multiplex degree of a node $i$ is a
  vectorial rather than a scalar quantity, a fact that allows to
  construct several degree-based biasing functions. In the following
we present two particular classes of such biasing functions, which we
call \textit{extensive} and \textit{intensive} biases, respectively.

\textit{Extensive bias functions. ---}
We call \textit{extensive} those walks whose motion rule depends on a
function of the degrees of the destination node at each of the $M$
layers. A first example is that of \textit{additive} degree-biased
walks, defined by transition probabilities of the form:
\begin{equation}
  \pi_{ji} \propto \sum_{\alpha=1}^{M} (k\lay{\alpha}_j)^{b_{\alpha}}
  \label{eq:additive}
\end{equation}
where $b_{\alpha}\in \mathbb{R}$ is the bias exponent associated to
layer $\alpha$. Another example is that of \textit{multiplicative}
degree-biased walks, whose transition probability takes the form:
\begin{equation}
  \pi_{ji} \propto \prod_{\alpha=1}^{M} (k\lay{\alpha}_j)^{b_{\alpha}}.
  \label{eq:multiplicative}
\end{equation}

We named these walks ``extensive'' since the number of free parameters
in the motion rule, namely the exponents $b_{\alpha}$, increases with
the number of layers $M$. This peculiar property of extensive walks
allows for a fine-grained setting of the bias in order to avoid nodes
whose replicas on each of the $M$ layers belong to a specific degree
class. For instance, in the case of a two-layer multiplex, if we set
$b_1 >0$ and $b_2 <0$ then the walkers will preferentially move
towards node having, at the same time, high degree on layer $1$ and
low degree on layer $2$. It might sometimes be desirable for a
  walker to have such sophisticated motion rules. An example is that
  of multiplex collaboration networks, in which nodes are scientists
  and layers represent co-authorship patterns in different fields. In
  that case, we might use an appropriately biased multiplex random
  walk which prefers to move towards nodes having a higher degree in a
  particular field, whose stationary probability distribution will
  represent a measure of the relative importance of each author in
  that field.

However, having a number of parameters which scales with the number of
layers is not always a desirable property, especially if one wants to
tune these parameters in order to obtain a walk with certain dynamical
properties (e.g., either in terms of stationary probability or in
terms of entropy rate). This problem is efficiently solved by
intensive bias functions.

\begin{figure*}[t]
  \begin{center}
      \includegraphics[width=6.0in]{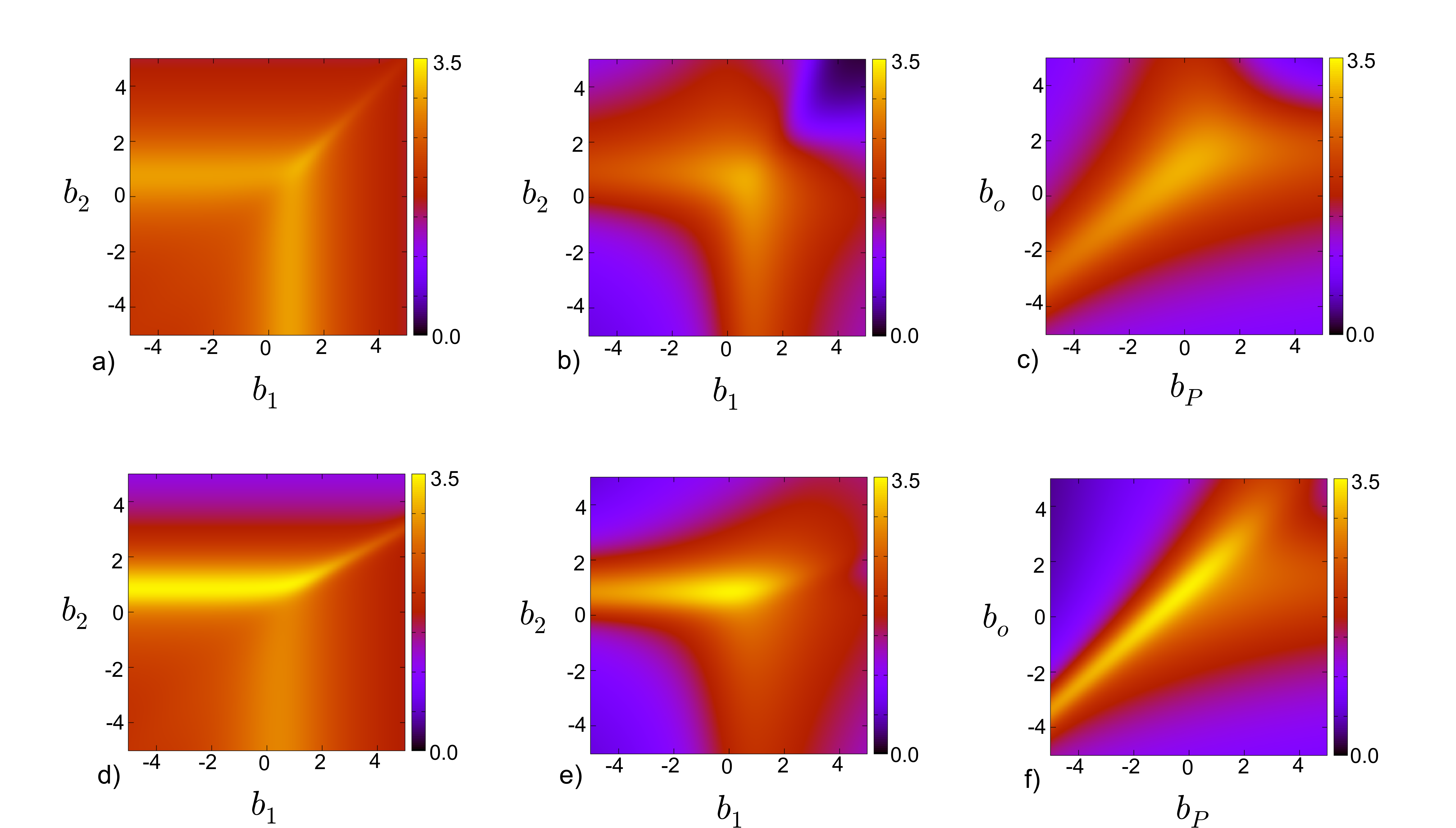}
  \end{center}
  \caption{Heat-maps of the value of entropy rate $h$ of
      different multiplex biased walks as a function of the parameters
      of the biasing function. The panels correspond, respectively, to
      additive [right, (a) and (d)], multiplicative [middle, (b) and
        (e)] and intensive walks [left, (c) and (f)] on uncorrelated
      duplex networks (in the top panels the two layers have the same
      power-law degree distribution $P(k) \sim k^{-\gamma}$ with
      $\gamma=2.5$, while in the bottom panels the two layers have
      power-law degree distributions with different exponents, namely
      $\gamma_1=2.2$ and $\gamma_2=2.7$. In general, the maximum of
      $h$ is obtained for positive values of the two bias parameters,
      corresponding to extensive walks which move preferentially
      towards nodes having high degrees on both layers, and to
      intensive walks whose motion rule is biased towards truly
      multiplex nodes.}
  \label{fig:fig1}
\end{figure*}

\textit{Intensive bias functions. ---} We call \textit{intensive}
those multiplex walks whose motion bias depends on one or more
intrinsically multiplex properties of the destination node. In the
following we will focus on the intensive walk whose transition
probability reads:
\begin{equation}
  \pi_{ji} = \frac{o_{ij}(o_{j}^{b_o}\mathcal P_{j}^{b_p})}{\sum_{\ell}
    o_{i\ell}(o_{\ell}^{b_o}\mathcal P_{\ell}^{b_p})}
  \label{eq:intensive}
\end{equation}
where $o_j=\sum_{\alpha}k\lay{\alpha}_j$ is the overlapping degree of
node $j$ and $\mathcal P_{j}$ is the multiplex participation
coefficient of $j$, and is defined as~\cite{Battiston2014}:
\begin{equation}
\mathcal P_i=\frac{M}{M-1}\left[1-
  \sum_{\alpha=1}^M\biggl(\frac{k_i^{[\alpha]}}{o_i}\biggr)^2\right].
\label{participationcoefficient}
\end{equation}
We notice that by considering $o_{\bullet}$ and $\mathcal P_{\bullet}$
we are effectively using information about the \textit{distribution}
of the edges of the destination node across the layers. In particular,
for fixed number of layers $M$, $o_i$ is proportional to the average
of the distribution defined by $\bm{k}_i = \{k\lay{1}_i, k\lay{2}_i,
\ldots, k\lay{M}_i\}$, while $\mathcal P_i$ gives information about
the homogeneity of $\bm{k}_i$, with $\mathcal P_i\sim 1$ if
$k\lay{\alpha}_i\simeq \frac{1}{M} \sum_{\beta}k\lay{\beta}_i \>
\forall \alpha$ (i.e., if node $i$ has roughly the same degree at all
layers) and $\mathcal P_i \sim 0$ if almost all the edges of node $i$
lie on just one layer.

We notice that when $b_o>0$ the walkers will preferentially move
towards hubs, while for $b_o<0$ they tend to visit the poorly
connected nodes more often.
Similarly, for positive values of $b_p$ the walkers will
preferentially move towards \textit{truly multiplex nodes}, i.e. nodes
whose distribution of edges across the $M$ layers is more homogeneous,
while for $b_p<0$ the walkers prefer to move towards \textit{focused
  nodes}, i.e. those having the majority of their connections in just
one or a few of the $M$ layers~\cite{Battiston2014}. In general, by
tuning the two parameters $b_o$ and $b_p$ we can obtain a rich variety
of different walks. For instance, for $b_o>0$ and $b_p>0$, the walkers
will be attracted by truly multiplex hubs (i.e., nodes with many
links, almost equally distributed across the layers). Conversely, when
$b_o>0$ and $b_p<0$ focused hubs are visited often and multiplex
poorly connected nodes are strongly avoided, and so forth. The
unbiased multiplex walk is recovered for $b_p = b_o = 0$.

The most interesting characteristic of the intensive walk defined by
Eq.~(\ref{eq:intensive}) is that the number of free parameters is
fixed and does not scale with the number of layers, as instead happens
for extensive walks. We will show in the following that intensive
walks usually perform at least as well as extensive walks, e.g. with
respect to the maximisation of entropy rate or to the minimisation of
heterogeneity in the stationary occupation probability distribution.

It is worth noting that in the case of a duplex, i.e. when
  $M=2$, even if the number of biasing parameters in intensive and
  extensive walks is the same, their effect on the motion of the
  walkers is different. Differently from $b_1$ and $b_2$, intensive
  biases do not allow to bias the walkers towards nodes with given
  properties in a particular layer but always consider intrinsically
  multiplex features, such as their total number of connections and
  their heterogeneity.

\begin{figure*}[t]
  \begin{center}
      \includegraphics[width=6.0in]{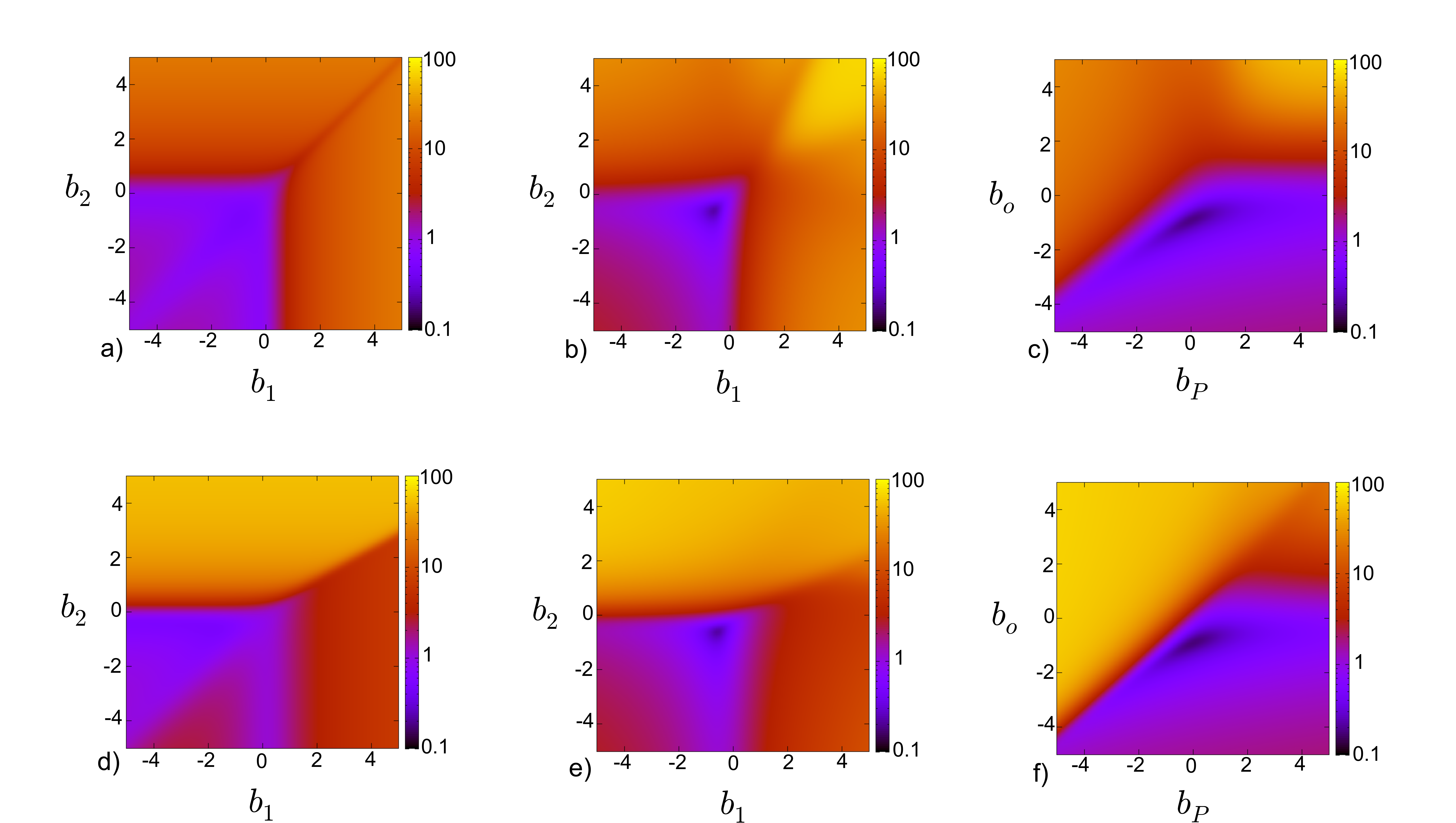}
  \end{center}
  \caption{ Heat-maps of the normalised standard deviation of
      the stationary occupation probability distribution $\eta(p^*)$
      of different multiplex biased random walks. Legend as in
      Fig.~\ref{fig:fig1}. In extensive walks, the minimum of $\eta$
      is always attained for negative values of the two exponents,
      while in intensive walks the minimum of $\eta$ is obtained for
      $b_o<0$ and $b_P\simeq 0$, meaning that walkers tend to
      preferentially move towards nodes with small degrees on both
      layers.}
  \label{fig:fig2}
\end{figure*}

In order to explore the differences in the dynamical
  properties (i.e., the entropy rate $h$ and the normalised standard
  deviation of the stationary occupation probability distribution
  $\eta(p^{*})$) of biased multiplex walks, in the top panel of
  Fig.~\ref{fig:fig1} we report the values of $h$ obtained by
  additive, multiplicative and intensive random walks as a function of
  the two bias exponents in a two-layer multiplex network whose layers
  have the same average degree $\langle k \rangle$ and power-law
  degree distributions $P(k)\sim k^{-\gamma}$ with $\gamma=2.5$, with
  no inter-layer correlations and no edge overlap\footnote{The results
    obtained for different values of the exponent $\gamma$ of the
    power-law degree distribution are comparable to those shown in
    Fig.~\ref{fig:fig1} and Fig.~\ref{fig:fig2}, and are not reported
    for brevity.}. We notice that also in this simple case the three
  walks have remarkably different behaviours. In particular, the
  additive walk exhibits a relative small sensitivity to the values of
  the biasing exponents, which results in smaller variations of
  $h$. In fact, there is a large region of $b_1$ (i.e. $0<b_1<2$)
  within which the entropy rate is almost constant and not very
  different from the absolute maximum for a relatively large range of
  values of the other exponent $b_2$, i.e. $-5<b_2<2$ (the same
  reasoning is valid for $0<b_2<2$ and $-5<b_1<2$, due to the symmetry
  of the additive bias function).

Conversely, the picture is much richer and less trivial in the
  case of multiplicative and intensive walks, for which the maximum of
  $h$ is obtained for a relatively small range of parameters, usually
  corresponding to positive exponents. We obtain slightly different
  results when we consider two layers with different power-law degree
  distributions $P(k\lay{1})\sim \left(k\lay{1}\right)^{\gamma_1}$ and
  $P(k\lay{2})\sim \left(k\lay{2}\right)^{\gamma_2}$, namely with
  exponents $\gamma_1=2.2$ and $\gamma_2=2.7$ respectively. In this
  case, the symmetry in the additive and multiplicative phase diagrams
  is broken, and the maximum values of $h$ are found by biasing the
  walk towards nodes with high degree on both layers, with a higher
  biasing exponent on the degree of the second layer, which has a more
  homogeneous degree distribution.  Also the phase diagram for the
  intensive walk is modified, with the line of maximum values becoming
  thinner.

Similar considerations hold for the phase diagram of $\eta(p^*)$,
reported in Fig.~\ref{fig:fig2}. In this case, the minimum variance
(yielding a more homogeneous exploration of nodes) is obtained for
negative values of the two bias exponents. Moreover, the phase diagram
exhibits quite small variations in the case of additive walk, while we
observe more heterogeneity in the case of multiplicative and intensive
walks. Again, the symmetry of the phase diagrams of the extensive
walks is broken when pairs of layers with different power-law
exponents $\gamma_1$, $\gamma_2$ are considered, with the region
$b_2>b_1$ showing greater variations than for $b_2<b_1$.
Qualitatively similar differences can be obtained with asymmetric
layers with respect to other statistical properties, such as density.

All the results for synthetic networks, both in the current and
following sections, have been obtained for layers with $N=10^4$ nodes
and averaged over $1000$ realisations.

\section{How the structure of a multiplex affects the walk}

In this section we illustrate how the structure of the multiplex
network affects the maximal entropy rate and the minimum
  heterogeneity of the stationary occupation probability distribution
  achievable in the system.

We focus on five structural aspects, namely \textit{i)} the
  presence and sign of inter-layer degree-degree correlations,
  \textit{ii)} the existence of edge overlap across layers,
  \textit{iii)} the number $M$ of layers of the multiplex,
  \textit{iv)} the power-law exponent $\gamma$ of the degree
  distribution of the layers, and \textit{v)} their density, measured
  through the average degree $\langle k \rangle$. Since our focus is
  on the construction of efficient walks (in terms of maximal
  dispersiveness and of homogeneity of the stationary occupation
  probability) the parameters of interest in all the cases are the
  overall maximum value of entropy rate, denoted by ${h}_{\rm max}$,
  and the minimum value of the normalised standard deviation, denoted by
  $\eta_{\rm min}$, obtained by extensive and intensive biased random
  walks as a function of the biasing parameters.

\begin{figure*}[!ht]
  \begin{center}
    \includegraphics[width=6in]{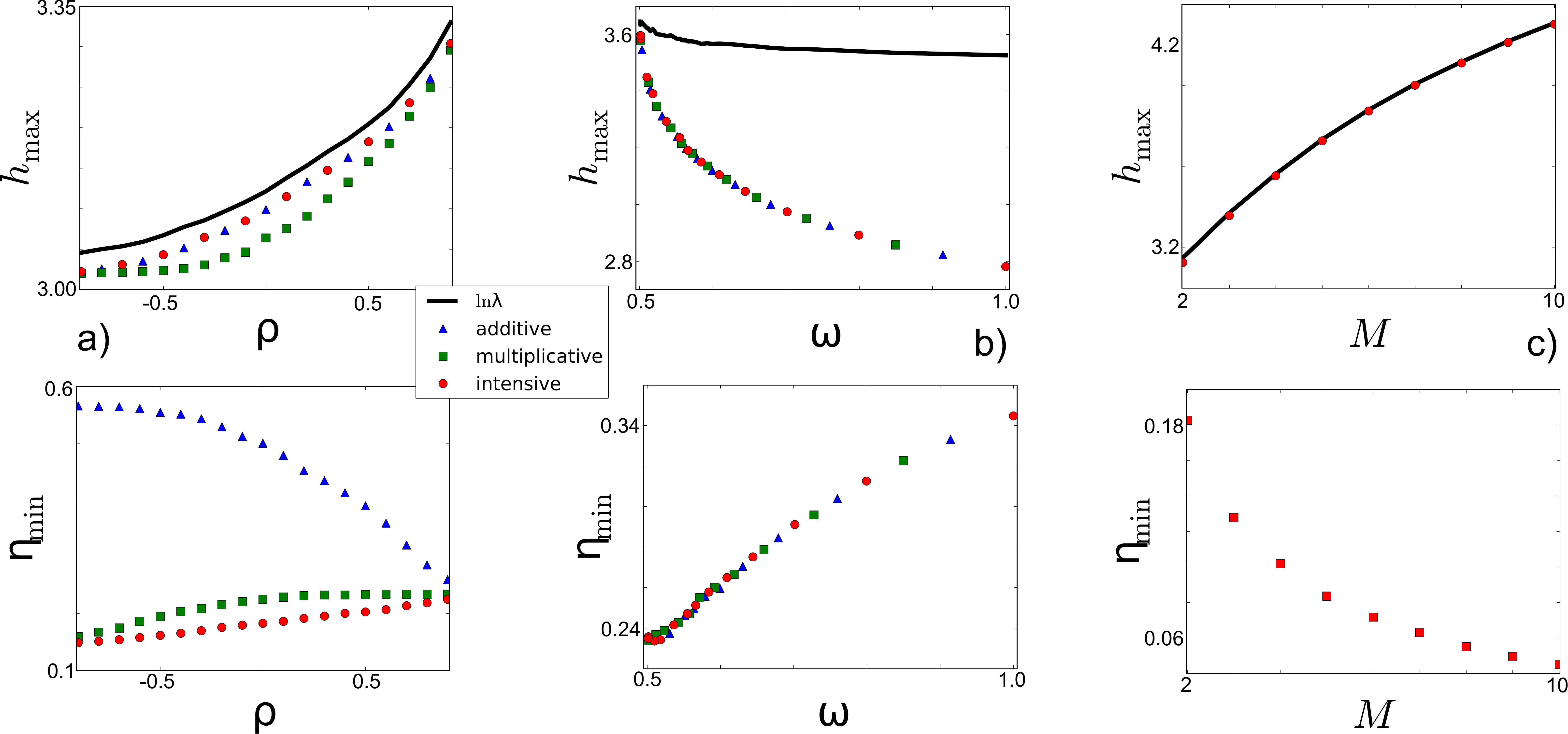}
  \end{center}
  \caption{Values of $h_{\rm max}$ (top panels) and $\eta_{\rm min}$
    (bottom panels) as a function of the the inter-layer degree
    correlation coefficient $\rho$ (a), the average edge overlap
    $\omega$ (b) and the number of layers $M$ (c), respectively for
    additive (triangles), multiplicative (squares) and intensive
    (circles) walks. For the entropy rate, we also show the value of
    $\widetilde{h}_{\rm max}=\ln \lambda$ corresponding to the maximum
    entropy random walk (solid line). (a) For all walks, ${h}_{\rm
      max}$ is an increasing function of the inter-layer degree
    correlation coefficient $\rho$, and provides a very good
    approximation of the maximum theoretical entropy rate
    $\widetilde{h}_{\rm max}$. Notice that intensive walks perform at
    least as well as the extensive ones. (b) As the overlap increases,
    the estimates of $h_{\rm max}$ obtained by the biased walks become
    less precise, while $\eta_{\rm min}$ increases as a function of
    $\omega$. (c) $h_{\rm max}$ increases and $\eta_{\rm min}$
    decreases as a function of $M$. In this case we only performed
    simulations for intensive walks.}
  \label{fig:fig3}
\end{figure*}

\textit{Effect of inter-layer degree correlations. ---} In a recent
work~\cite{Nicosia2014corr} the authors have shown that real-world
multiplex networks are usually characterised by non-trivial
inter-layer degree correlation patterns. In the same paper the authors
propose several methods to quantify the presence of inter-layer
correlations between a pair of layers, including the rank correlation
among the two degree sequences, as measured by the Spearman's
coefficient $\rho$.  If we call $R\lay{\alpha}_i$ the rank of node $i$
due to its degree on layer $\alpha$, the Spearman rank correlation
coefficient between layer $\alpha$ and layer $\beta$ reads:
\begin{equation}
  \rho_{\alpha, \beta} = \frac{\sum_{i}\left(   R\lay{\alpha}_i -
        \overline{ R\lay{\alpha} }    \right) 
\left( R\lay{\beta}_i -
       \overline{ R\lay{\beta} }      \right)}
    {\sqrt{\sum_i\left(   R\lay{\alpha}_i -
        \overline{ R\lay{\alpha} }    \right)^2\sum_j \left(  
 R\lay{\beta}_j -  \overline{ R\lay{\beta} } 
\right)^2}}
\end{equation}
where $\overline{ R\lay{\alpha} }$ and $\overline{ R\lay{\beta} }$ are
the average ranks of nodes respectively at layer $\alpha$ and layer
$\beta$. The coefficient $\rho$ takes values in $[-1,1]$, so that
$\rho=1$ if the two degree sequences are perfectly correlated (meaning
that a hub at layer $\alpha$ is also a hub at layer $\beta$), while
$\rho=-1$ when the two degree sequence are perfectly anti-correlated,
i.e. when a hub on layer $\alpha$ is always a poorly connected node on
the other layer, and viceversa. Intermediate positive
(resp. negative) values of $\rho$ indicate weaker positive (negative)
inter-layer correlations, while $\rho\simeq 0$ when the two degree
sequences are uncorrelated.

In Fig.~\ref{fig:fig3}(a) we report the plot of $h_{\rm max}$
  and $\eta_{\rm min}$ for extensive and intensive walks on two-layer
  multiplex networks with same average degree and power-law degree
  distributions $P(k)\sim k^{-\gamma}$ with $\gamma=2.5$, for
  different levels of inter-layer degree correlations. As made
evident by the figure, intensive walks usually perform at least as
well as extensive walks with respect to both maximisation of entropy
and minimisation of the heterogeneity of the stationary occupation
probability distribution. This suggests that, aside from the actual
differences in the phase space, intensive walks are able to span the
same range of values of entropy and $\eta(p^*)$ by using only two
parameters, irrespective of the actual numbers of layers in the
multiplex.

\textit{Effect of edge overlap. ---}
We now investigate the impact of the presence of edge overlap on the
long-term dynamics of extensive and intensive walks. We recall here
the definition of overlap for an edge $(i,j)$, which is the fraction
of layers in which the edge $(i,j)$
exists~\cite{Battiston2014, Bianconi2013}, i.e.:
\begin{equation}
  \omega_{ij} = \frac{1}{M}\sum_{\alpha =1}^{M} a\lay{\alpha}_{ij}.
\end{equation}
The edge overlap of a multi-layer network is defined as the average of
$\omega_{ij}$ over all the node pairs for which $o_{ij}\neq 0$ (i.e.,
for all pairs of nodes which are connected by at least one edge):
\begin{equation}
  \omega=\frac{1}{\sum\limits_{i,j} (1 - \delta_{o_{ij},
      0})}\sum_{ij}\omega_{ij}=\frac{1}{2K}\sum_{ij}\omega_{ij}
\end{equation}
where $K$ is the number of pairs of nodes which are connected
  in at least one of the $M$ layers. Notice that the average edge
  overlap ${\omega}$ is equal to $1$ only if all the $M$ layers are
  identical, while ${\omega}=1/M$ when every edge is present on
  exactly one of the $M$ layers.

\begin{figure}[!ht]
  \begin{center}
    \includegraphics[width=3in]{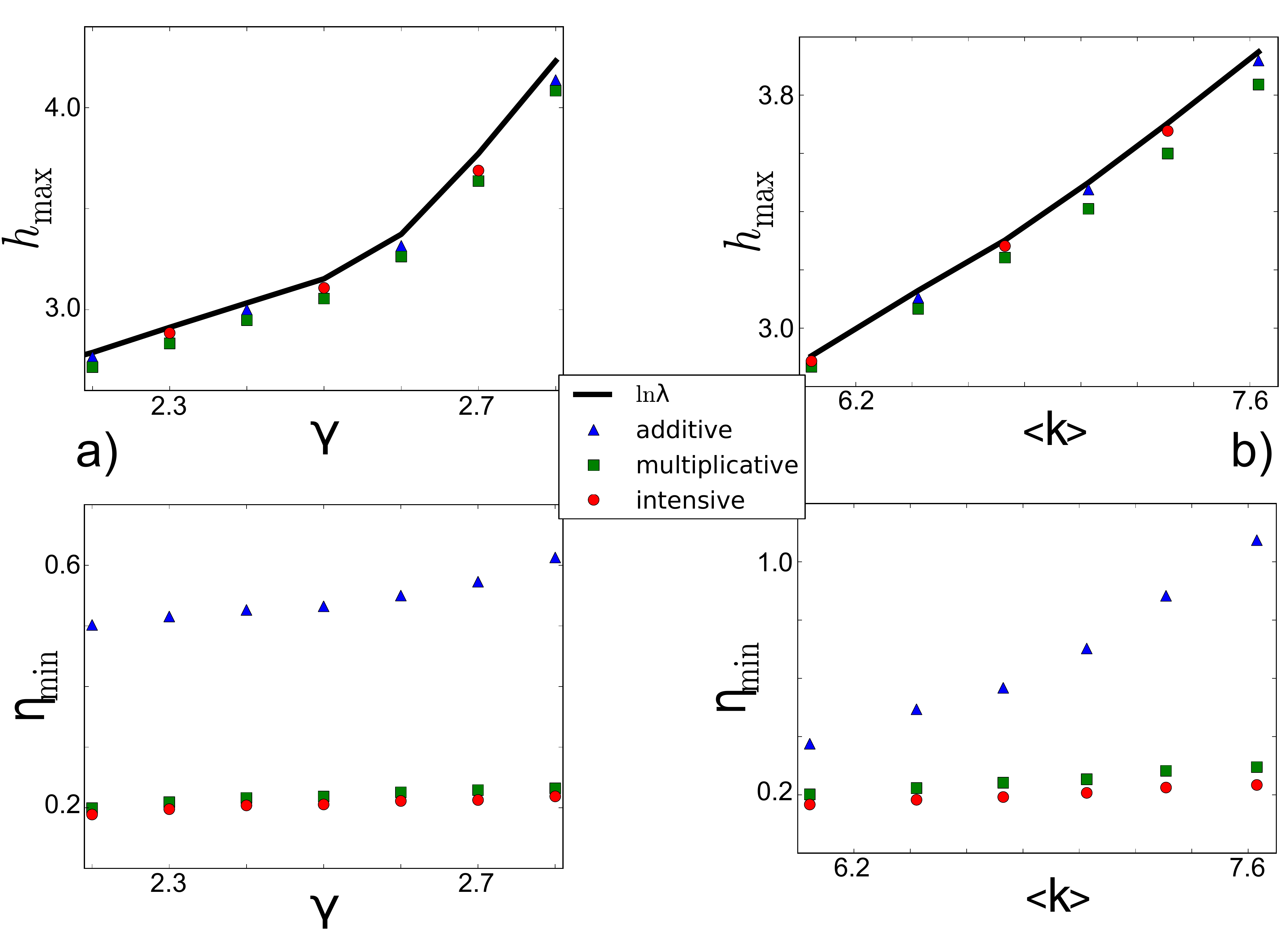}
  \end{center}
  \caption{Values of $h_{\rm max}$ (top panels) and $\eta_{\rm
        min}$ (bottom panels) as a function of the exponent $\gamma$
      of the the power-law distribution of each layer (a) and of the
      average degree $\langle k \rangle$ (b), respectively for
      additive (triangles), multiplicative (squares) and intensive
      (circles) walks. For the entropy rate, we also show the value of
      $\widetilde{h}_{\rm max}=\ln \lambda$ corresponding to the
      maximum entropy random walk (solid line). As shown, for all
      walks $h_{\rm max}$ appears to increase as a function of both
      $\gamma $ and $\langle k \rangle$. Smaller variations are also
      found in the values of $\eta_{\rm min}$.}
  \label{fig:fig4}
\end{figure}

We started from two-layer multiplex networks obtained by
  coupling identical layers (thus having edge overlap equal to $1$)
  with power-law degree distributions $P(k)\sim k^{-\gamma}$ with
  $\gamma=2.5$, and then we obtained multiplex networks with
  prescribed values of edge overlap by rewiring a certain percentage
  of the edges of one of the two layers in order to maintain the
  degree sequence unaltered. Notice that by construction the resulting
  multiplex networks have maximally positive inter-layer degree
  correlations (i.e., $\rho=1$). As shown in Fig.~\ref{fig:fig3}(b),
  $\eta_{\rm min}$ becomes higher as $\omega$ increases, meaning that
  higher values of edge overlap correspond to a more heterogeneous
  stationary state probability distribution. Conversely, $h_{\rm max}$
  decreases with $\omega$, in accordance with the fact that higher
  edge overlap tends to hinder the dispersiveness of the walk, since a
  smaller number of distinct walks can originate from each
  node. Summing up, multiplex networks having smaller values of edge
  overlap are overall preferable in order to maximise the
  dispersiveness of the walk and to obtain a more homogeneous
  stationary occupation probability. In other words, a small edge
  overlap guarantees a more effective exploration of a multiplex
  network and, at the same time, a more homogeneous distribution of
  the probability of visiting each node.

\textit{Effect of the number of layers. ---} It is also interesting to
study how the dynamical properties of intensive walks change when the
number of layers $M$ is progressively increased. To this aim, we
constructed multiplex networks with different number of layers, with
no inter-layer degree correlations and negligible edge overlap, where
all the layers had power-law degree distributions $P(k)\sim
k^{-\gamma}$ with $\gamma=2.5$.  As shown in
  Fig.~\ref{fig:fig3}(c), $h_{\rm max}$ is an increasing function of
  $M$, while $\eta_{\rm min}$ decreases as the number of layers
  grows. In general, the addition of layers in absence of inter-layer
  correlation flattens the structural differences among the nodes of
  the multiplex, and provides better dispersiveness and less
  heterogeneity in the occupation probability distribution.

\textit{Effect of the heterogeneity of the degree distribution. ---}
We investigate here how the heterogeneity of the the degree
  distribution of each layer affects $h_{\rm max}$ and $\eta_{\rm
    min}$.  To this aim, we considered pairs of uncorrelated layers
  with the same power-law degree distribution $P(k)\sim k^{-\gamma}$
  for different values of $\gamma$, maintaining fixed the average
  degree of the networks $\langle k \rangle$.  The plots in
  Fig.~\ref{fig:fig4}(a) confirm that both $h_{\rm max}$ and
  $\eta_{\rm min}$ grow as $\gamma$ increases, i.e. as the degree
  distribution of the layers becomes more homogeneous. We notice
  though that the variation in $\eta_{\rm min}$ appears to be
  relatively smaller, especially for multiplicative and intensive
  walks. This result can be explained by considering that
  dispersiveness is favoured by more homogeneous degree
  distributions. Layers with different power-law exponents $\gamma_1$
  and $\gamma_2$ have been considered in the previous section.

\textit{Effect of layer density. ---} Finally, we focus on the
  effect of layer density, measured through the average degree of the
  layers $\langle k \rangle$. Once again we report here the case of
  uncorrelated layers with power-law exponent $\gamma=2.5$, but
  similar results have been obtained for other values of $\gamma$. As
  shown in Fig.~\ref{fig:fig4}(b), both $h_{\rm max}$ and $\eta_{\rm
    min}$ increase as a function of $\avg{k}$. Layers with different
  average degrees $\langle k\lay 1 \rangle$ and $\langle k\lay 2
  \rangle$ break the symmetry of the phase diagrams for $h$ and $\eta$
  qualitatively in a similar way as pairing layers with different
  power-law exponents.

Summing up, the analysis of the impact of structural
  properties on the values of $h_{\rm max}$ and $\eta_{\rm min}$
  attainable on a multiplex network confirms that positive inter-layer
  degree correlations, small edge overlap, large number of layers, and
  more homogeneous layers all concur towards allowing biased random
  walks with nearly-optimal dispersiveness and closely-to-homogeneous
  steady-state visiting probability. In other words, a multiplex
  network with a large number of layers and small edge overlap, where
  nodes have roughly the same number of links at all layers, can be
  explored ways more efficiently than a similar multiplex network
  where nodes have disassortative inter-layer correlations and edges
  are redundant across layers.
  
  In the following section we show that the multiplex airline
  transportation networks of all the six continents have evolved
  towards a structure which provides a good trade-off between
  efficient exploration and robustness.

\section{Applications to real-world airline transportation networks}
As an application, we study here the dynamical properties of
  multiplex biased walks on a set of real-world systems, namely the
  six continental airline transportation networks.  In such systems
  nodes represent airports, edges indicate the existence of a route
  between two airports and each layer is associated to an airline
  company, i.e. all the edges in a layer represent the routes operated
  by the corresponding airline. Such networks have been introduced and
  extensively studied in Ref.~\cite{Nicosia2014corr}.  As shown in
  Table~\ref{tab:1}, all such multiplex networks consist of a
  relatively high number of layers. For this reason, we will use
  intensive walks to compute the maximal entropy rate $h_{\rm max}$
  and the minimum value of the standard deviation of the stationary
  distribution $\eta_{\rm min}$. In Table~\ref{tab:1}, we also report
  for each multiplex the average number of layers $M \times \omega$
  where each edge exists, the theoretical upper value of entropy rate
  $\ln \lambda$, and the values of $h_{\rm max}$ and $\eta_{\rm min}$
  obtained by optimising intensive walks.

\begin{table}[ht]
\centering
\begin{tabular}{|l||l|l||l||l|l||l|}
\hline
Multiplex &  $N$ & $M$ & $M \times \omega$ & $\ln \lambda$ & $h_{\rm max}$ & $\eta_{\rm min}$\\
\hline
Africa & 238 & 84 & 1.57 & 3.36& 2.20 & 1.36\\
Asia & 795 & 213 & 2.16 & 4.96   & 3.52 & 1.17\\
Europe & 594 & 174 & 1.55 & 4.60 &  3.76 & 1.06\\
North America & 1029 & 143 & 1.56 & 4.70 & 3.75 & 1.35\\
Oceania & 261 & 37 & 1.52 & 3.71 & 2.39 & 2.00\\
South America & 300 & 58 & 1.81 & 3.66 & 2.59 & 1.08\\
\hline
\end{tabular}
\caption{\label{tab:1} Structural properties of the six
    continental airline transportation systems. For each multiplex, we
    report the number of nodes $N$, the number of layers $M$, the
    average number of layers in which an edge exists $M \times
    \omega$, the theoretical upper value of entropy rate $\ln
    \lambda_{\rm max}$ and the extremal values $h_{\rm max}$ and
    $\eta_{\rm min}$ obtained by optimising intensive walks.}
\end{table}

We notice that the efficiency of a transportation system is usually
measured in terms of the accessibility of the locations it serves. In
particular, in an ideal transportation system it should be easy to
travel between any pair of far-apart regions of the network, mostly
irrespective of where exactly those locations are located. Now,
discarding the cost associated to the distance between the nodes of an
airline transportation network, high accessibility can be obtained by
guaranteeing that a traveller can reach remote locations in the system
without large effort, in terms of number of interchanges, and that all
locations can be visited with comparable effort. We have seen that in
the language of random walks these two criteria correspond,
respectively, to the maximisation of dispersiveness and to the
minimisation of the standard deviation of the visiting probability.

Hence, we can ask whether the six continental air transportation
systems can guarantee a good level of navigability, i.e. an optimal
trade-off between dispersiveness and homogeneity of the visiting
probability. We reckon that a more informative analysis of the
efficiency of these systems would require more detailed information
about the actual patterns of trips travelled by passengers, the cost
associated to each route, the presence of non-Markovian effects
(people often come back to their original place at the end of a trip),
the non-stationarity of the system due to seasonality, etc.. However,
we argue that biased random walks can still provide useful, yet
coarse-grained, information about the overall navigability of those
systems. Since we cannot modify the degree distributions of each of
the layers, or the patterns of inter-layer correlations, or the actual
number of layers in each continental air transportation system, we
focus here in particular on the effects of edge overlap.

In the previous section we showed that networks with high edge overlap
$\omega$ achieve lower maximal values of dispersiveness of the walk
and larger heterogeneity of the equilibrium occupation probability
distribution. When two nodes are connected by more than one edge,
indeed, from a dynamical point of the view some connections are
wasted, since redundant links do not allow for new paths in the
network. However, their redundancy might often be important for a
transportation system, since it makes specific connections more robust
to single link failures. It is not unrealistic to assume that
multi-layer transportation systems from the real-world have developed
by satisfying a trade-off between the necessity to provide, at the
same time, high diffusivity together with reasonable levels of
robustness.  

Because of the large heterogeneity in the size and number of layers of
the six continental transportation systems, it is necessary to
introduce some kind of normalisation which allow to compare the
results observed in different systems. In order to test the effect of
edge overlap on the diffusion properties of real-world systems, for
each of the six multiplex networks we computed the z-score of the
average edge overlap:
\begin{equation}
 z(\omega) = \frac{\omega-\langle \omega \rangle}{\sigma(\omega)},
\end{equation}
where $\langle \omega \rangle$ and $\sigma(\omega)$ represent
respectively the average value and the standard deviation of the
overlap computed on an ensemble of suitably randomised multiplex
networks. In particular, for each continental airline system we
sampled $1000$ multi-layer graphs from the configuration model which
maintains fixed the degree sequence of all the layers and rearranges
the links on each layer, pairing edge stubs at random. We computed
also $z(h_{\rm max})$ and $z(\eta_{\rm min})$, i.e. the z-scores of
the maximal entropy rate and minimum variance over those $1000$
multiplex graphs.

\begin{figure*}[t]
  \begin{center}
      \includegraphics[width=6.0in]{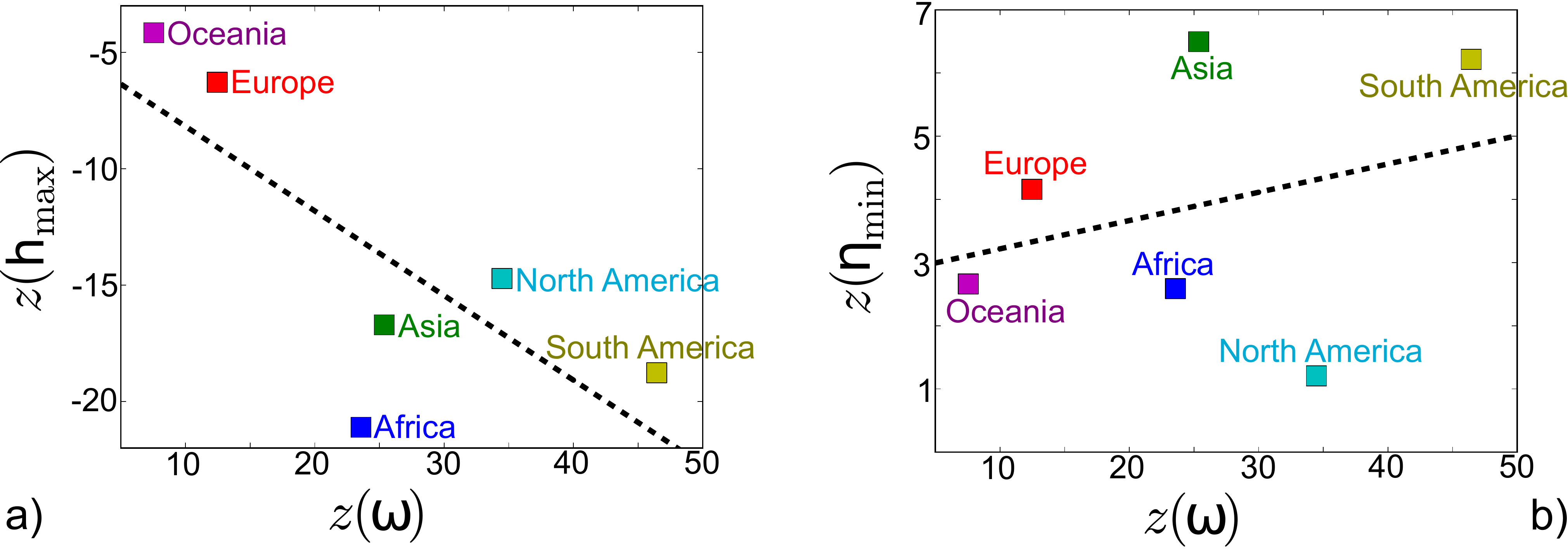}
  \end{center}
  \caption{z-score of the average edge overlap $\omega$ versus
      the z-scores of the maximal entropy rate $h_{\rm max}$ (a) and
      the minimum standard deviation of the stationary distribution
      $\eta_{\rm min}$ (b) obtained through intensive walks. In
      agreement with findings for synthetic networks, also in
      real-world systems $z(h_{\rm max})$ is negatively correlated
      with $z(\omega)$ - Pearson's correlation coefficient $r=-0.70$ -, whereas $z(\eta_{\rm min})$ is positively
      correlated - $r=0.30$, which increases to $r=0.67$ excluding the outliner North America -.}
  \label{fig:fig5}
\end{figure*}

The results reported in Fig.~\ref{fig:fig5} confirm that also in
real-world systems $h_{\rm max}$ is negatively correlated with edge
overlap, in agreement with the results obtained on synthetic
networks. Similarly, $\eta_{\rm max}$ is positively correlated with
$\omega$.  Notice that we have $z(\omega)>0$ in all the six
continents, meaning that the edge overlap of the real-world systems is
always higher than that of the null-model, in agreement with the
observation that real-world transportation networks tend to guarantee
a certain level of robustness to failures. However, the quest for
robustness has a cost in terms of dispersiveness and accessibility. In
fact, $h_{\rm max}$ is consistently smaller than the value observed in
the randomised systems ($z(h_{\rm max})<0$) for all continents, and
similarly the steady-state probability distribution is consistently
larger than that observed in the null model ($z(\eta_{\rm min})>0$)

It is quite interesting to note that the two multiplex networks with
smallest overlap and overall better diffusion properties are the
continental networks of Oceania and Europe, which span the least
geographical space. We can speculate that in such systems some nodes
representing cities in different countries are connected comparably
well by different modes of transport, such as trains and bus, suited
for relatively short distances and not included in our analysis. This
might potentially explain the relative low number of redundant edges
in those two airline transportation systems. Conversely, the necessity
to provide route redundancy has somehow forced the airline
transportation networks of Asia, South America and North America
towards slightly less efficient configurations.

\section{Conclusions}
In our work we have explored how to extend biased random walks to the
case of multiplex networks, showing that the richness of multi-layer
systems allows to define several different classes of walks.  In
particular we studied the general features of the so-called extensive
walkers (where the node properties, as the degree, are considered
separately at each of the layers with different biasing parameters)
and intensive walkers (biased on of the product two intrinsically
multiplex, namely the overlapping degree and the participation
coefficient) finding closed forms for the stationary occupation
probability of these walks and for the entropy rate, and provided
simplified heterogeneous mean-field expressions for the case in which
the multiplex has no correlations.

We thoroughly investigated how structural properties of the multiplex,
such as its number of layers, the presence of edge overlap and/or
inter-layer degree correlations, the density of the layers and
  the heterogeneity of their degree distribution affect the dynamics
of the random walkers. We found that number of layers, edge overlap
and inter-layer degree correlations have a substantial impact on the
diffusion properties of the walks. Also, we found that intensive
random walkers perform at least as well as extensive random walkers in
all the considered scenarios, with the advantage that the number of
bias parameters does not scale with the number of layers.

Finally, the study of the diffusion properties of six
  real-world multiplex networks, namely the continental airline
  transportation networks of Africa, Asia, Europe, Oceania, North and
  South America, has shed some new light on the interplay between
  efficiency and robustness in multi-layer transportation systems. In
  particular, we found that the emerging necessity to provide some
  resilience to single link failures, which corresponds to the
  introduction of some level of edge overlap, has shaped these systems
  in such a way that their navigability, in terms of entropy rate and
  heterogeneity of the node occupation probability, has somehow been
  sacrificed in favour of robustness. The results of the present work
  represent a valuable theoretical contribution to the development of
  efficient strategies to explore, search or navigate multiplex
  networks, and confirm the importance of appropriately taking into
  account the multiplexity of interactions when modelling
  intrinsically multi-dimensional systems.

\section{acknowledgments}
F.B., V.N., and V.L. acknowledge the support of the EU Project LASAGNE,
Contract no. 318132 (STREP). V.L. also acknowledges support from EPSRC project GALE Grant EP/K020633/1.
This research utilized Queen Mary's MidPlus computational facilities,
supported by QMUL Research-IT and funded by EPSRC grant EP/K000128/1.

\end{document}